# Non-volatile Electric Field Control of Thermal Magnons in the Absence of an Applied Magnetic Field

*Eric Parsonnet[1], Lucas Caretta[2], Vikram Nagarajan[1], Hongrui Zhang[2], Hossein Taghinejad[1], Piush Behera[2,4], Xiaoxi Huang[2], Pravin Kavle[2], Abel Fernandez[2], Dmitri Nikonov[3], Hai Li[3], Ian Young[3], James Analytis[1], Ramamoorthy Ramesh[1,2,4]*

1. *Department of Physics, University of California, Berkeley, California 94720, USA*
2. *Department of Materials Science and Engineering, University of California, Berkeley, California 94720, USA*
3. *Components Research, Intel Corporation, Hillsboro, Oregon 97124, USA*
4. *Material Science Division, Lawrence Berkeley National Laboratory, Berkeley, California 94720, USA*

## Abstract:

Spin transport through magnetic insulators has been demonstrated in a variety of materials and is an emerging pathway for next-generation spin-based computing. To modulate spin transport in these systems, one typically applies a sufficiently strong magnetic field to allow for deterministic control of magnetic order. Here, we make use of the well-known multiferroic magnetoelectric, $BiFeO_3$, to demonstrate non-volatile, hysteretic, electric-field control of thermally excited magnon current in the absence of an applied magnetic field. These findings are an important step toward magnon-based devices, where electric-field-only control is highly desirable.

## Main:

In the field of magnonics, spin waves, rather than electrons, form the fundamental operating unit [1–4]. The field has experienced rapid growth over the last decade as exciting progress has yielded a breadth of interesting physics as well as the potential for low power dissipation in computing [5]. In lieu of an electronic current, insulating magnetic materials can host magnon currents, which carry spin information and avoid Ohmic losses associated with electron transport. Such materials are well-suited for wave-based computing based on magnon logic [6–10]. Indeed, theoretical work has proposed antiferromagnetic spin wave field-effect transistors [11] and realizations of all-magnon transistors based on magnon-magnon scattering with resonant excitation have already been experimentally demonstrated [9]. There are several ways to create magnons [1,4,12–14], and spin transport via magnon currents have already been reported in a variety of magnetic systems [5,15–19]. Though resonant excitations are typically used to study spin waves [20,21], magnon currents can be excited incoherently by a thermal gradient through the spin Seebeck effect (SSE) [22]. While other near-DC-frequency incoherent excitation mechanisms exist [12,23], thermal excitation of magnons is better suited to materials with complex domain structure since it does not require long-range magnetic order [24]. Previous research has demonstrated non-local spin transport [25] in insulating ferrimagnets [12,26,27] and antiferromagnets [5,15,17–19,28], thermally excited spin-transport over exceptionally long distances [29,30], and non-volatile magnetic field control [19]. Electric field control of magnon spin transport, however, has been limited to concurrent application of high magnetic fields [24]. For operational devices based on magnon transport, electric field control in the absence of an applied magnetic field could be a crucial advance for the field.

Here, we make use of the well-known multiferroic material, $BiFeO_3$ (BFO), to demonstrate such electric field control of thermal magnons. BFO is a room-temperature magnetoelectric with a large ferroelectric polarization (~90 μC/cm^2), G-type antiferromagnetic ordering, and a weak ferromagnetic moment arising from the Dzyaloshinskii-Moriya interaction [31–33]. The ferroelectric and antiferromagnetic domain structures in BFO exhibit a one-to-one correspondence [34], and deterministic control of magnetic order via manipulation of the ferroelectric state (with applied electric fields) has already been demonstrated [35–37], making BFO an attractive option for high-speed, low energy computing [38–41]. Previous work on BFO [42,43] has revealed broad tunability of the magnon dispersion with applied electric field, and early theoretical work predicted all electrical switching of magnon propagation [44].

In this letter, we demonstrate a novel manifestation of magnetoelectric coupling in BFO to manipulate magnon current. Magnons are excited via the SSE and spin transport is detected via the inverse spin Hall effect (ISHE) [45,46]. We demonstrate non-volatile, hysteretic, bistable states of magnon current and establish a robust means of switching between the two states with an applied electric field. Via piezo-response force microscopy (PFM), we reveal the switching pathway of the ferroelectric order, which is accompanied by the switching of net canted magnetic moment, providing the mechanism for electric field control of magnon current.

We grow BFO samples via pulsed laser deposition (**Supplementary Material, Methods**) and employ a non-local device geometry (**Fig. 1a**) consisting of two lithographically defined (**Supplementary Material, Methods**) Pt wires separated by a distance $d$ ($\leq 1$μm) on the BFO surface. In the channel between the Pt electrodes, we observe well-ordered $109^\circ$ ferroelectric domains (**Fig. 1a**). This confirms the high quality of the BFO film and, via the established correspondence between ferroelectric and magnetic order, allows us access to the magnetic domain structure of the device [34,37,47]. Each of the four leads (**Fig. 1a**) is wire-bonded for the non-local measurement, described next. In the "measurement configuration," low-frequency AC current, $f = 7$ Hz, driven through the injector Pt wire (**Fig. 1b**) creates a thermal gradient in the film via Joule heating of the Pt (heater) wire. This thermal gradient excites magnons via the SSE. The resulting magnon current is detected at the detector Pt wire as a voltage along the length of the wire, originating from ISHE and spin scattering at the Pt/BFO interface [12,46,48]. We use the standard lock-in technique to measure the magnon current as a differential voltage along the detector wire, $V_{nl}$. Since the thermal magnon current scales with the square of the charge current in the injector wire, we measure the second harmonic of the detected voltage, *i.e.* $V_{nl}(2\omega)$, (**Supplementary Material, Methods**).

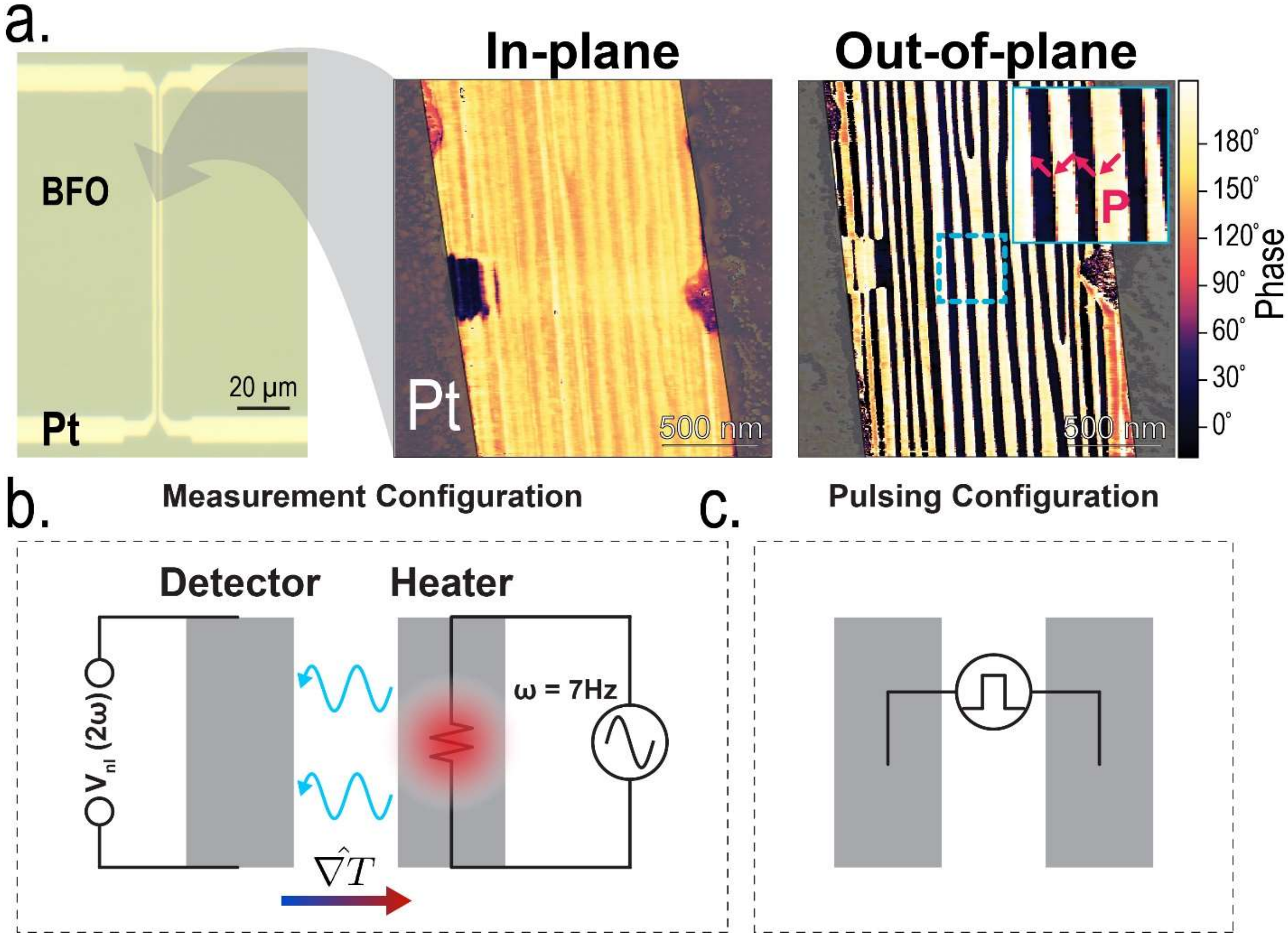

***Figure 1. Experimental Setup***. ***a.*** *Optical and PFM images of non-local device structure. Out-of-plane (OOP) and in-plane (IP) PFM images reveal a well-ordered 109° domain structure. Arrows in inset (OOP) show IP projection of spontaneous polarization,* ***P***. ***b.*** *Measurement configuration.* ***c.*** *Pulsing configuration.*

We confirm the efficacy of our device structure and non-local ("measurement configuration", **Fig. 1a**) testing protocol by fabricating identical devices on $Y_3Fe_5O_{12}$ (YIG) (**Supplementary Material, Methods**) and performing the prototypical in-plane magnetic-field dependent nonlocal measurement (**Supplementary Material Fig. S1**), which shows the expected behavior [12]. To modulate the magnon current in BFO in the absence of an applied magnetic field, we perform *in-situ* electric-field pulsing across the channel ("pulsing configuration", **Fig. 1c**) thereby switching the ferroelectric, and consequently magnetic order parameters.

We follow an experimental protocol (**Fig. 2b**) designed to both confirm the switching of the ferroelectricity and monitor resulting changes in detected magnon current. Following a unipolar (5ms, 350 kV/cm) voltage pulse in the pulsing configuration, we measure the second harmonic voltage on the detector wire, $V_{nl}(2\omega)$, in the measurement configuration. We then confirm the ferroelectric state by measuring a single bipolar ferroelectric hysteresis (PE) loop. As observed, (**Fig. 2b**), the PE loop shows only one switching event (*e.g.,* only showing switching in the positive direction, following negative poling) confirming that the ferroelectric state is both switched and remnant. We then switch the pulse polarity and start again. By alternating positive and negative polarity electric field pulses (**Fig. 2c**) one can clearly observe two non-volatile states of measured magnon current.

Owing to the ISHE detection mechanism, the Pt detector wire acts as a directional detector, sensitive to the component of incident magnon spin polarization orthogonal to the length of the wire (**Fig. 2a**) [5,12,19]. The existence of two states of magnon current, therefore, indicates that the electric field induced switching results in changes to the magnon spin polarization pointing across the channel. Our results thus indicate that the switching of the ferroelectricity induces switching of the net canted moment (**M**) adjacent to the detector Pt wire. As a result, the spin polarization of thermally excited magnons along **M**, also flips, resulting in the observed change in detected non-local voltage. To better understand the mechanism of such switching, we turn to PFM to directly image the ferroic order.

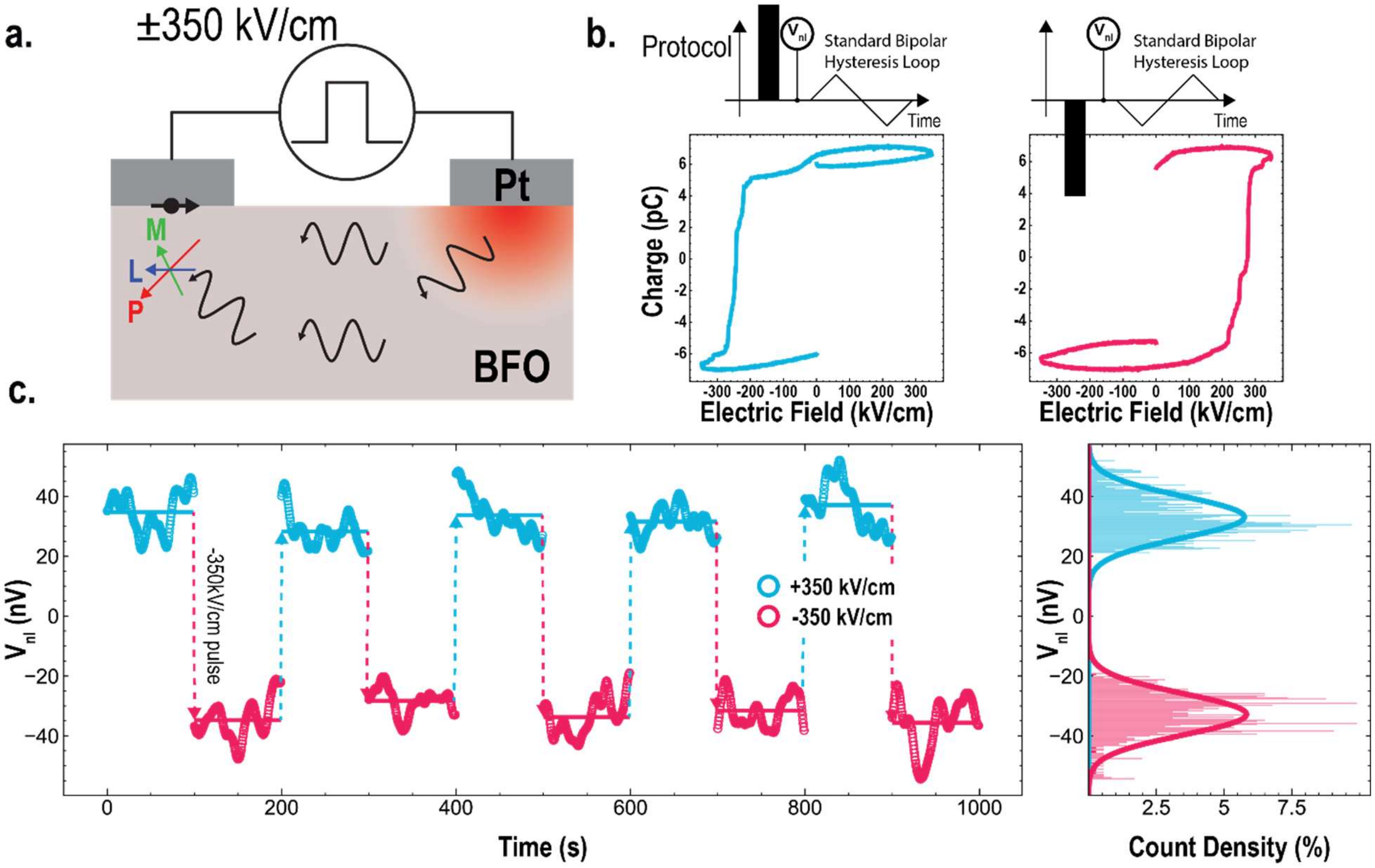

***Figure 2. Bi-stable states of thermal magnon current**. **a.** Cross sectional device schematic. As shown by the interfacial spin (black), the detected voltage along the detector (left) wire is dependent on the magnon spin polarization component orthogonal to the length of the Pt wires. **b.** Experimental protocol and results of "half" hysteresis loops confirming the stability of the ferroelectric state after electric field poling. **c.** Measured lock-in second harmonic voltage,* $V_{nl}(2\omega)$*, measuring magnon current, as a function of time. 100 seconds of data are collected after each electric-field pulse. Data*

*reflects relative changes upon poling, i.e., a small (~10s of nV) DC offset is subtracted from both positively and negatively poled signals. Histogram combining results from 10 trials confirms two distinct states of magnon current. Fits are to normal distributions. An 800 µA charge current was used to generate the thermal gradient for SSE.*

It has been shown previously [49], that there is one-to-one correspondence between ferroelectric domains and antiferromagnetic domains in BFO, so, via PFM we are able to intuit the magnetic domain structure. Within a single domain, the ferroelectric polarization, Néel vector and canted ferromagnetic moment are oriented orthogonal to one another (**Fig. 3a**) with the in-plane projection of the canted moment pointing along the in-plane projection of the ferroelectric polarization [35,50,51]. Owing to mechanical and electrostatic boundary conditions [47], adjacent domains' polarization vectors are oriented $109^\circ$ from each other and aligned head to tail (**Fig. 3a**). The existence of such $109^\circ$ domains is confirmed via PFM in our films (**Fig 1a.**), and results in a net canted moment (along $<100>_{pc}$), which points orthogonal to the directionality ($<010>_{pc}$) of the stripe domains (**Fig. 3**). To study the switching of the ferroic state, we perform PFM imaging after application of $\pm$350 kV/cm across the channel. We observe reversal of the in-plane contrast (**Fig. 3b**), indicating in-plane reversal of the ferroelectric polarization. Importantly, while the in-plane component of the polarization reverses, the underlying ferroelastic domain structure is preserved, *i.e.*, switching occurs within each ferroelastic (stripe) domain. The persistence of the ferroelastic domains very likely contributes to the reversibility of the magnonic signal observed upon bipolar electric field pulsing (**Fig. 2c**). The IP PFM results, in combination with out-of-plane (OOP) PFM imaging after switching (**Supplementary Material Fig. S2**), which does not show reversal, allow us to conclude that polarization switching proceeds via a $71^\circ$ IP switch, consistent with previous research [35]. We show, schematically (**Fig. 3c**), the resulting reorientation of the polarization, Néel vector, and canted moment after $71^\circ$ switching. The ISHE detection mechanism is sensitive to the magnon spin polarization, and therefore the magnetic order, directly beneath the detector Pt wire. Furthermore, since the Pt wire spans several domains, the detected voltage is a function of the adjacent *net* magnetic order (*i.e.* the sum of twin domain contributions). While magnons have been shown to traverse both ***M*** and ***L*** [12,15,19], from the schematic (**Fig. 3c**), one can observe that while the net canted moment, **M**, does reverse, the net Néel vector, **L**, does not reverse following $71^\circ$ switching. This indicates that our data is sensitive to spin wave excitations carried along (spin polarization antiparallel to) the net canted moment, **M**.

While we have so far discussed thermal magnons, it is important to address another excitation mechanism, namely the spin accumulation mechanism (SAM) [5,12,19,48] from the spin Hall effect (SHE) in the injector wire. This effect is dependent on the charge current in the injector wire (as opposed to $j^2$ in SSE) and therefore appears in the first harmonic nonlocal voltage. Electric-field switching of BFO results in no change in the first harmonic signal (**Supplementary Material Fig. S3**). It has been shown both experimentally [17] and theoretically [52,53] that (anti-) ferromagnetic domain walls act as scattering sites for incident magnons. To observe a first harmonic signal, magnons excited via SAM at the injector must traverse the channel without (or with minimal) scattering. Our PFM data reveal many domain walls between the injector and detector, which scatter all SAM excited magnons, and result in no observed first harmonic signal. Thermal magnons however, can be "re-excited" after a domain wall since the thermal gradient, governed by phonon diffusion, for example, persists. As described in detail in **Supplementary Material Section 1,2**, similar physics can explain the lack of first harmonic signal in multidomain NiO [14], and is corroborated by reports on YIG which study the effects of nonlocal thermal gradients [30] and heat-transparent/spin-opaque interfaces [54].

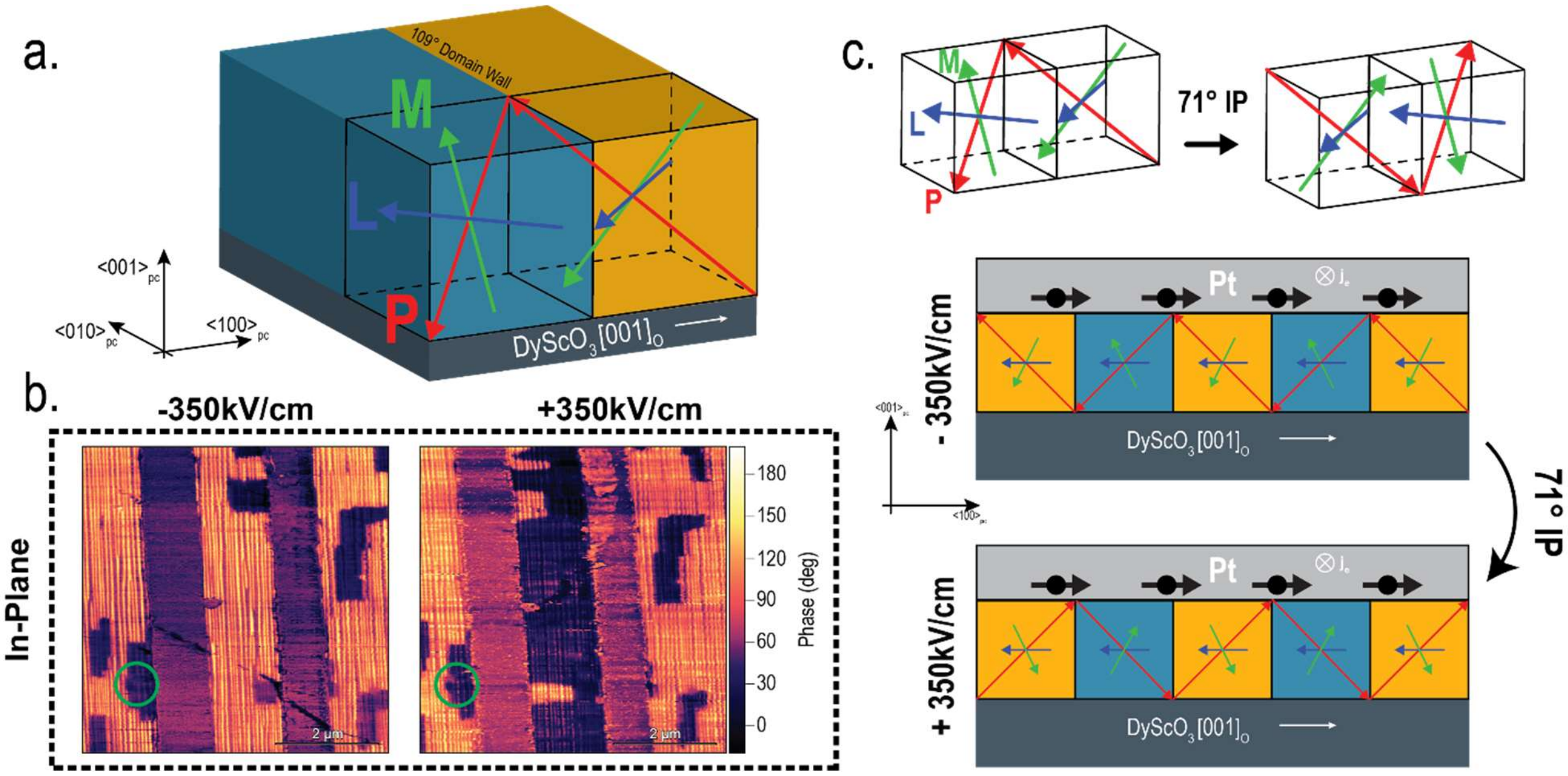

***Figure 3. Switching Mechanism. a.*** *Schematic of twin $109^{\circ}$ domains showing ferroelectric polarization vector,* ***P*** *(red), Néel vector* ***L*** *(blue), and canted magnetization vector* ***M*** *(green).* ***b.*** *In-plane phase PFM images after +350kV/cm and -350kV/cm applied across the channel. The change in contrast indicates reversal of the net in-plane polarization (and therefore canted magnetization). A (Green) circle marks an external reference domain pattern for comparison.* ***c.*** *Schematic of $71^{\circ}$IP switch showing reversal of the both the net ferroelectric polarization,* ***P****, and net canted moment,* ***M****.*

Having established the mechanism behind magnetization reversal and the observed magnon current, we now demonstrate its hysteretic nature. We perform a quasi-static measurement (**Fig. 4a**), varying the magnitude of the electric field pulse across the channel from negative to positive and back again, while measuring (over 100 seconds) the non-local signal after each applied electric field pulse. We observe a hysteretic response in the magnon current (**Fig. 4b**) which closely matches the ferroelectric hysteresis loop of the same device. To confirm that the observed data does not stem from capacitive charging or other extrinsic circuit effects [55], we perform the identical measurement on YIG (**Fig. 4b**). As expected, we observe no ferroelectric hysteresis, and importantly no hysteretic magnon current. Next, to confirm that the signal does not come from the remanent state of the ferroelectric polarization alone, we use a non-magnetic, in-plane ferroelectric $Pb_{0.7}Sr_{0.3}TiO_3$ (PSTO) (**Supplementary Material, Methods**) sample with a similar value of switchable charge and again perform the identical experiment (**Fig. 4b**). Here, we observe a strong ferroelectric hysteresis response, as expected, but do not observe any hysteresis in the nonlocal voltage. The YIG and PSTO control samples together, therefore, allow us to conclude that the BFO signal is magnetic in nature. Finally, as the SSE signal scales with the square of the charge current in the injector wire, we expect a linear dependence on injector (heater) power ($I^2R$) in the differential nonlocal voltage, defined as the difference between measured nonlocal second harmonic voltage when poled with a positive vs. negative electric field. We show (**Fig. 4c**) the expected linear dependence as a function of heater power.

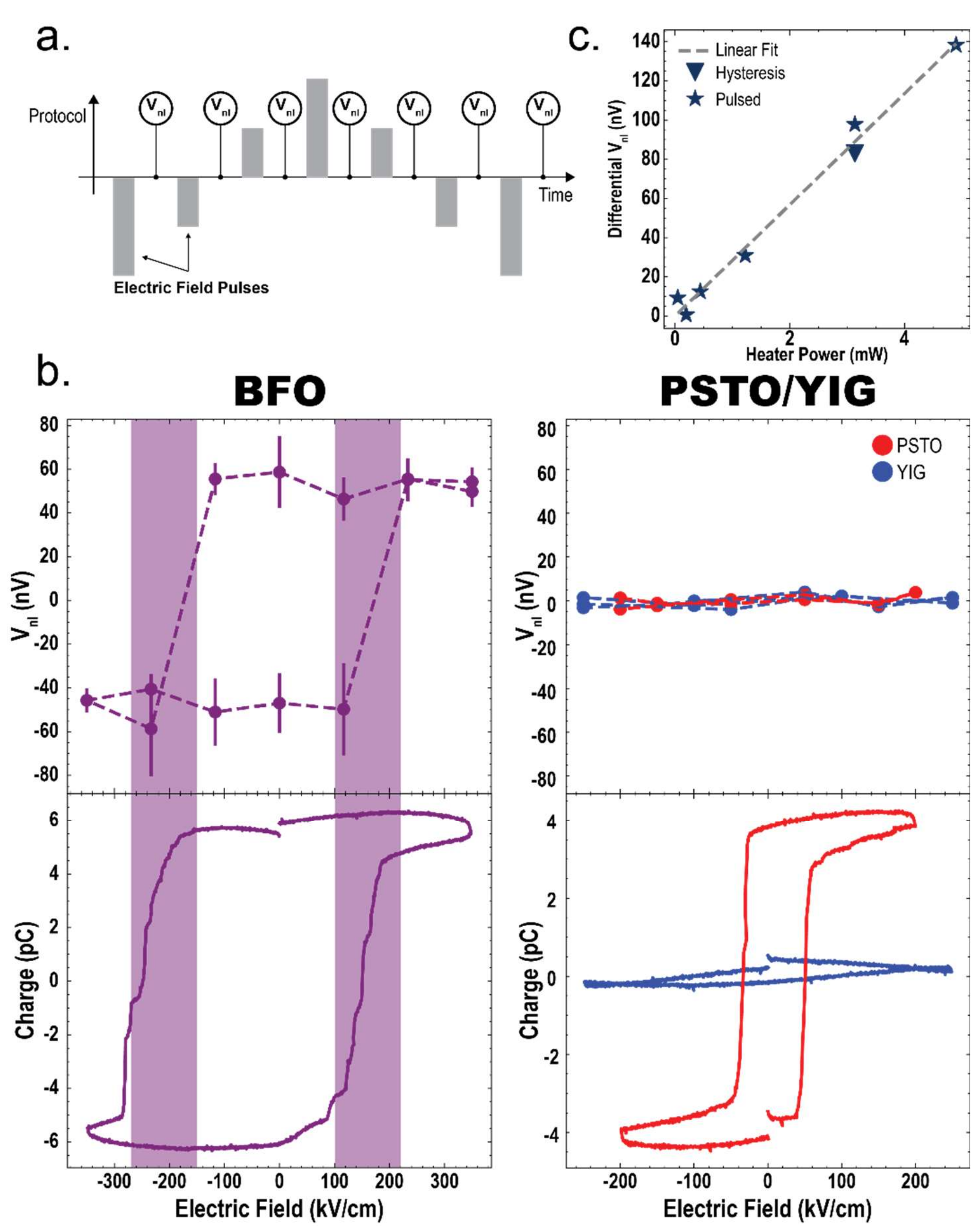

***Figure 4. Hysteretic Response. a.*** *Hysteretic magnon current measurement protocol.* ***b.*** *Observation of hysteresis in nonlocal second harmonic signal in BFO showing excellent agreement with the associated ferroelectric hysteresis loop. Identical measurements on PSTO and YIG are also shown.* ***c.*** *Magnitude of differential nonlocal voltage as a function of injector (heater) current, as measured through several different means.*

In conclusion, we have demonstrated a novel manifestation of intrinsic magnetoelectric coupling in BFO, establishing electric field control of non-volatile, hysteretic, bi-stable states of magnon current in the absence of an applied magnetic field. This represents a crucial step towards operational magnon-based devices. On-going work focuses on several pathways for increasing the magnitude of the non-local voltage for practical applications [40]. By varying the domain structure with choice of substrate [47], one can vary the number and type of magnon scattering sites present between the injector and detector. Advanced lithography techniques can also be utilized to minimize the injector-detector distance, with previous research indicating highly favorable scaling laws for reduced channel widths [12,29]. In fact, with improved magnon coherence and/or decreased channel spacing, domain walls can be written using an out-of-plane electric field (rather than in-plane as demonstrated here), thereby enabling a non-volatile three terminal transistor which operates on magnon scattering at domain walls at the gate. Perhaps most importantly, however, is the inclusion of alternate spin-orbit (SO) metals (replacing Pt). The spin Hall angle sets an intrinsic limit on the detected voltage, while the interface between the SO metal and the BFO can limit spin conductance and introduce variability in the fabrication process. Oxide SO metals, such as $SrIrO_3$, have recently shown high spin hall angles [56], and most importantly can be grown epitaxially, *in-situ*, via PLD on BFO, likely allowing for improved spin conductance, higher non-local voltages, and lower operating current, while maintaining BFO quality. The results shown here, offer an initial verification, highlighting an important synergy between multiferroics and magnonic spintronics, and demonstrating a novel pathway toward functional magnonic devices.

# Non-volatile Electric Field Control of Thermal Magnons in the Absence of an Applied Magnetic Field (Supplementary Material)

## Methods

*Film Growth (BFO)*

100nm thick $BiFeO_3/DyScO_3$ (BFO) films were grown via pulsed laser deposition at $690 - 710°C$ with focused laser fluence ~1.2 $J/cm^2$ under 100-160 mTorr oxygen pressure and cooled down to room temperature at 500 Torr oxygen pressure. These are established, well-characterized growth conditions which allow us to create model, epitaxial systems that are used for the studies presented here.

*Film Growth (YIG)*

$Y_3Fe_5O_{12}$ (YIG) films were grown on $Gd_3Ga_5O_{12}$ (0 0 1) substrates by RHEED assisted pulsed laser deposition. The YIG films were deposited at 700 °C and 100-mtorr oxygen partial pressure from the chemical stoichiometric ceramic target by using the KrF excimer laser (248 nm) at the energy density of 1.5 $J/cm_2$ and the repetition rate 5 Hz. After growth the samples were cooled down at a rate of 10 °C/min in atmospheric oxygen pressure.

*Film Growth (PSTO)*

100 nm thick $Pb_{0.7}Sr_{0.3}TiO_3/DyScO_3$ (110) (CrysTec GmbH) (PSTO) films were grown via pulsed laser deposition using a KrF excimer laser (248 nm, LPX 300, Coherent). The PSTO growth was carried out at a heater temperature of $640°C$ in a dynamic oxygen pressure of 100 mTorr with a laser fluence of 1.75 J/cm2. The laser repetition rates of 10 Hz and 2Hz were used for $Pb_{1.2}TiO_3$ ceramic target (Kurt J. Lesker) and $SrTiO_3$ single crystal target, respectively. The desired composition was achieved via sub-unit-cell-level material mixing utilizing the synchronized targets motor motion and laser pulse sequence. The 20% excess lead in $Pb_{1.2}TiO_3$ target was found to be vital to compensate the lead loss during growth. Following the growth, the sample was cooled to room temperature at $10°C$/min. in a static oxygen pressure of 700 Torr. The as-grown films were also characterized by X-ray diffraction including symmetric θ-2θ line scans and 2D reciprocal space mapping (RSM) studies using a high-resolution X-ray diffraction (XRD) system (X'pert Pro2, PANalytical). All films studied herein were found to be fully epitaxial and highly crystalline.

*Piezo-response Force Microscopy*

Making use of an atomic force microscope (MFP-3D, Asylum Research), we conducted dual a.c. resonance tracking PFM using a conductive Pt/Ir-coated probe tip (NanoSensor PPP-EFM) to image the in-plane and out-of-plane domain structures.

*UV Lithography + Pt Liftoff*

Devices (**Fig. 1**) are patterned via UV lithography using a Heidelberg MLA150 Maskless Aligner in the Berkeley Marvell NanoLab at CITRIS, using AZ MiR 701 Photoresist. After patterning a blanket layer of ~10nm of Pt is deposited via room temperature DC magnetron sputtering at 2 mTorr Ar pressure (base pressure of ~$10^{-8}$ Torr). Pt is then lifted off via ~5 hr soak in 1-methyl-2-pyrrolidone (NMP) at $85°$C.

*Thermal Magnon measurements*

Non-local second harmonic measurements are performed using a Keithley 6221 current source to run current through the heater wire, and two identical Stanford Research Systems Model SR830 Lock-in amplifiers are used to simultaneously measure the first and second harmonic differential voltage along the detector wire. All measurements, unless otherwise stated, are performed at a frequency of 7 Hz in the absence of an applied magnetic field.

The measured resistance across the channel, *i.e.*, from the heater to detector is >20GΩ. As measured by a Keithley 2400 multimeter, the resistance measures overload. These results hold both before and after electric field poling, and cycling between states. We are confident the measured signal does not stem from leakage currents between the electrodes.

*Data Acquisition/Analysis was performed using ekpy* [1]

**Figure S1. Non-local measurement on YIG. a.** Orientation of devices, relative to zero of magnetic field. A charge current is driven through the injector wire (labeled with $j_e(+)$ and $j_e(-)$) and the nonlocal voltage at the **b.** first harmonic (sensitive to the spin accumulation mechanism) and **c.** second harmonic (sensitive to the spin Seebeck effect) are measured in detector wire (labeled with $V_+$ and $V_-$). The charge current amplitude is 500μA. As expected, we observe a 90 ° phase shift in the observed nonlocal signal for devices (Case 1, 2) oriented orthogonal to each other.

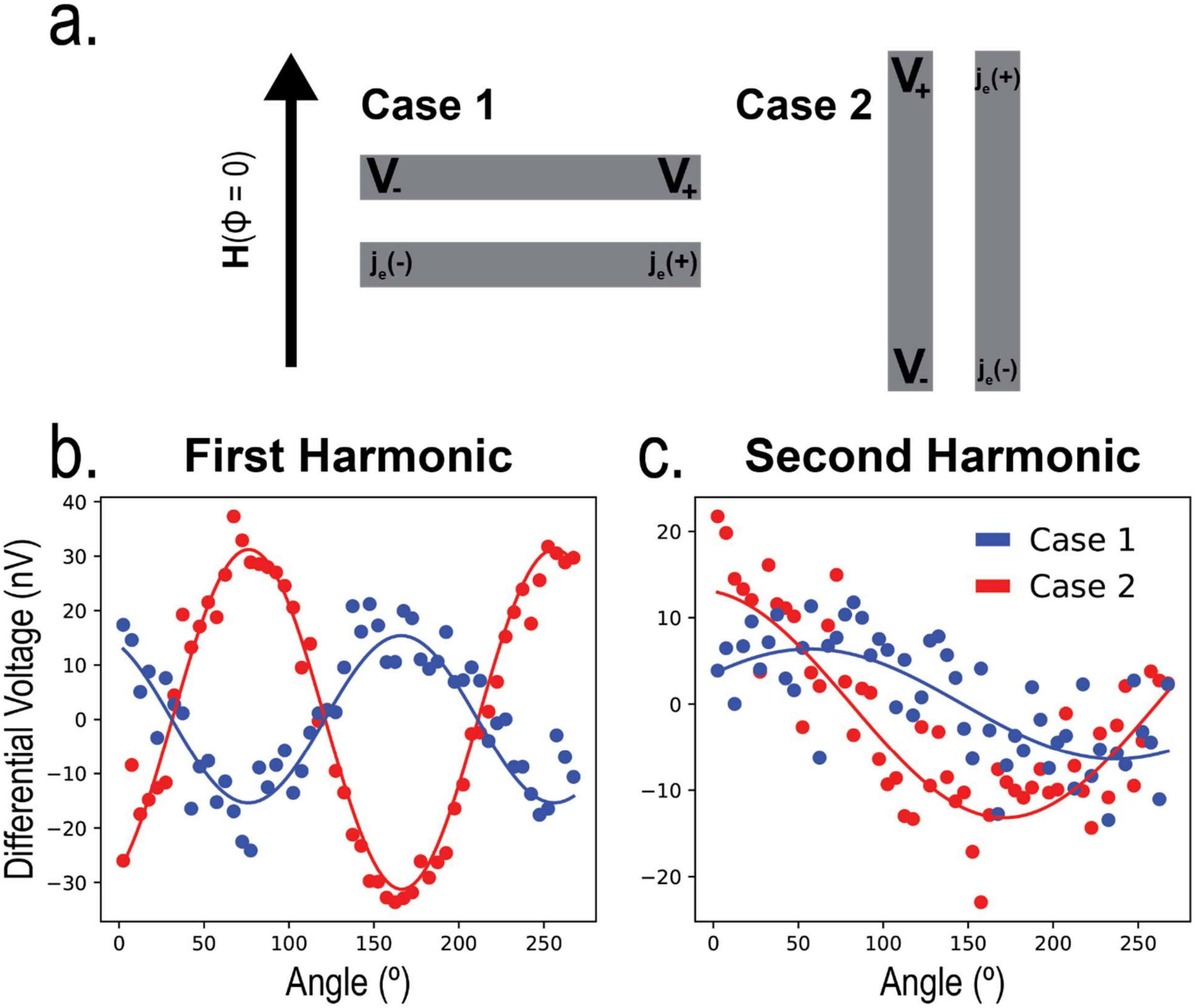

**Figure S2. Out-of-plane PFM after $\pm$350kV/cm poling**. Green circles highlight identical area outside of the channel as a reference state. One can readily observe no out-of-plane switching, evidencing that poling of the channel results in 71 degree switching of domains.

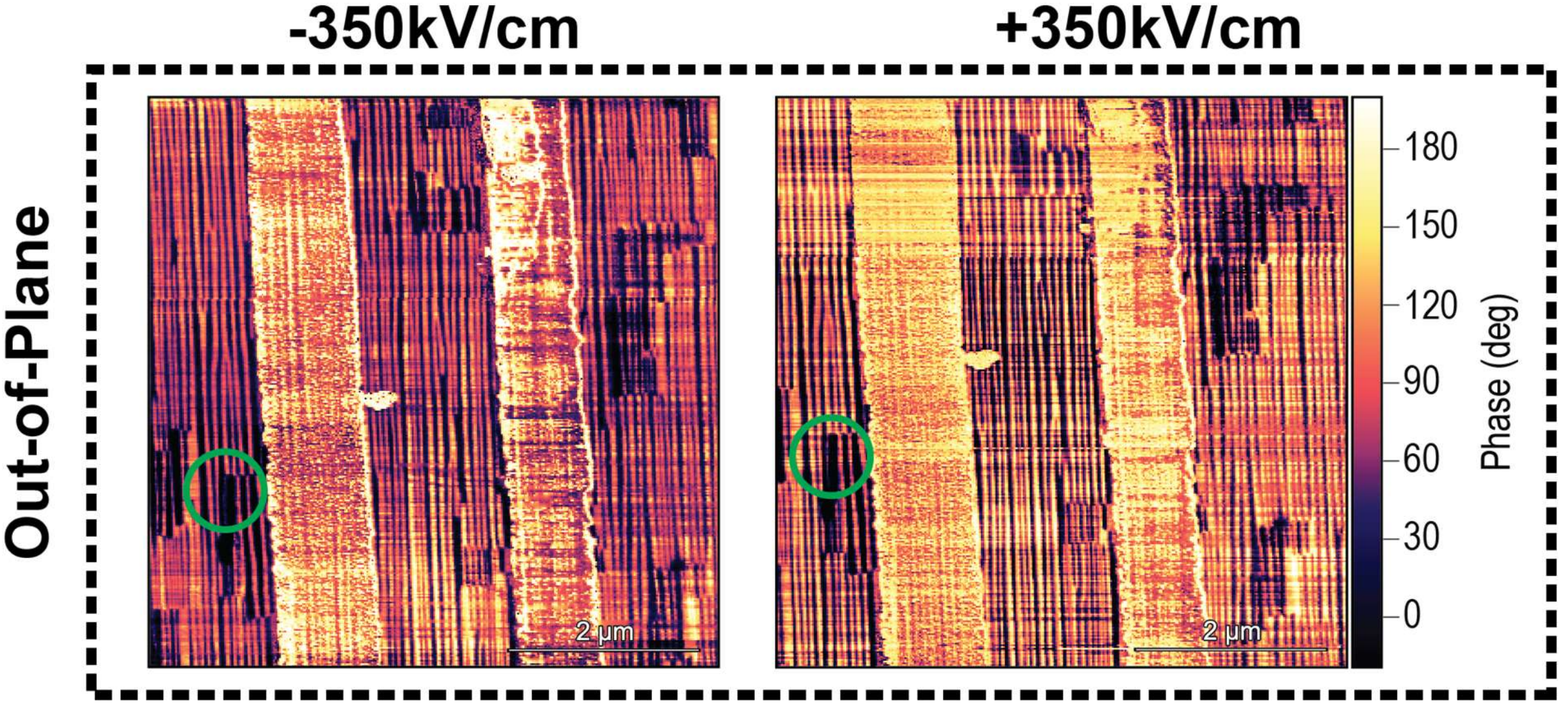

**Figure S3. First-Harmonic non-local voltage upon Electric-Field pulsing.** Average first harmonic signal before and after application of 350kV/cm electric field pulse, measured concurrently with second harmonic voltage (**Main Figure 2**). The data reveal no repeatable change in first harmonic signal before and after switching.

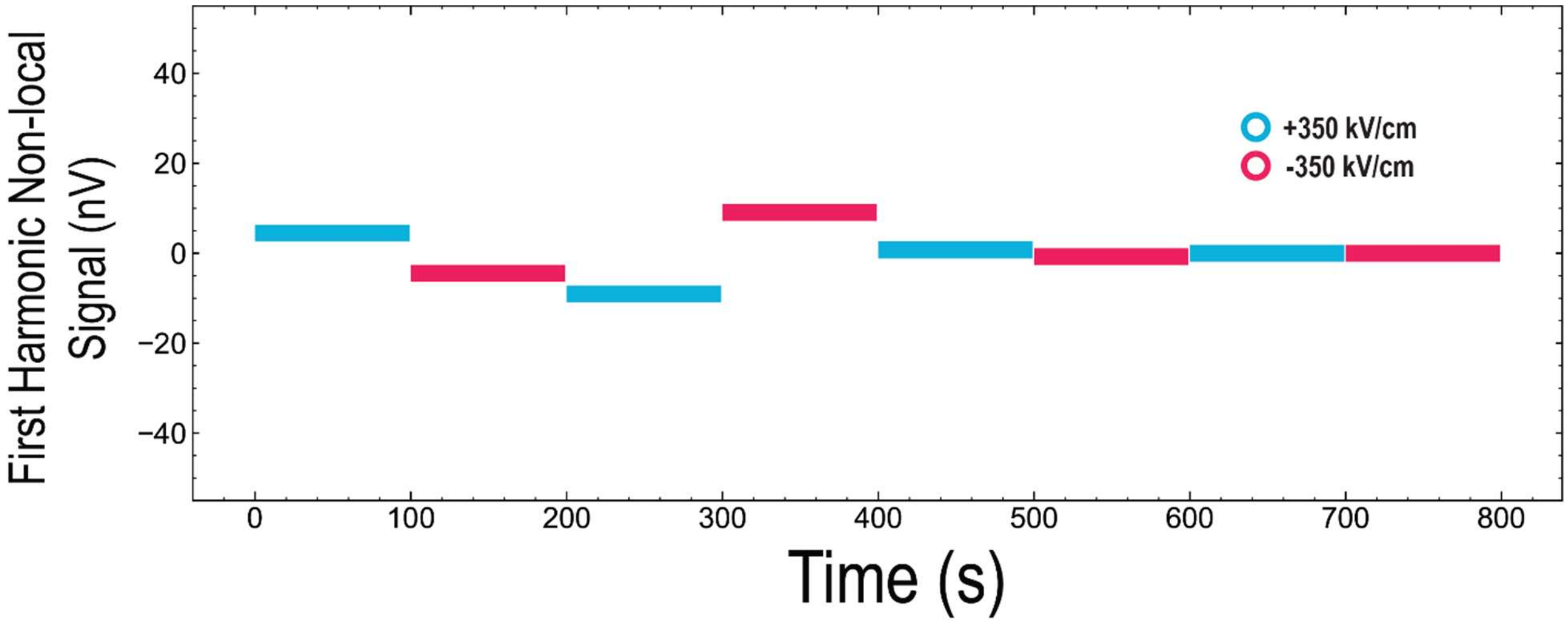

### Section 1. Extended discussion of first vs. second harmonic signal

The first harmonic and second harmonic signals, while related, probe slightly different physics. As described in [2], the first harmonic measures spin transport via magnons excited by the spin accumulation mechanism (often referred to as electrical excitation in the literature), while the second harmonic probes thermally excited magnons. In systems such as YIG, a ferrimagnetic insulator with low Gilbert damping and long-range magnetic order, the two processes are closely related and indeed, the authors in [2] find similar values of magnon spin diffusion lengths for both first and second harmonic signals. To measure a first harmonic signal, magnons, which are excited locally at the injector, must diffuse to the detector without scattering or significant attenuation. The authors treat the thermal excitation similarly, where Joule heating at the injector excites thermal magnons which then diffuse to the injector. However, later work [3] has shown that the picture of local Joule heating at the injector is incomplete, since the thermal gradient persist beyond the local region surrounding the injector, and the heating/spin-current source are delocalized.

Though the first and second harmonic signals exhibit similar length scalings in YIG, more complex systems such as multidomain NiO [4], where the authors observe a thermal signal, but no electrical signal, $Cr_2O_3$ [5], where researchers observe a thermal, but no electrical signal, or $Fe_2O_3$ [6], where the authors find a spin diffusion length of the electrical signal to be ~9 microns, but the thermal signal to persist beyond 80 microns, require a more nuanced investigation of the two signals. To understand why we observe similar effects in BFO, we must return to the origin of the two signals.

To observe the first harmonic signal, electrically excited magnons at the injector must diffuse to the detector without damping, scattering or encountering other attenuation mechanisms. It has been shown theoretically that magnetic domain walls act as scattering sites for incident magnons [7,8]. This has also been shown experimentally, in the same non-local geometry to that is implemented in our experiment, in $Fe_2O_3$ [9], where multidomain films exhibit significantly shorter magnon attenuation lengths, as measured by the spin accumulation mechanism (*i.e.,* first harmonic) signal. As more domains are introduced between the injector and detector the first harmonic signal will decrease owing to scattering of the magnons in the channel between the detector and injector. We see the exact same effect in BFO. In BFO, magnetic domains and ferroelectric domains are known to exhibit a 1-1 correspondence [10,11], so, by our piezo-response force microscopy imagining **(Main Text Fig. 1, 3)**, we are able to confirm the existence of many magnetic domain walls (which are pinned in place, owing to multiferroicity in BFO and the dominant ferroelectric and ferroelastic energy scales) that act as scattering sites for incident magnons. As demonstrated in [12], thermal gradients can persist through spin-opaque interfaces, meaning the thermal gradient can persist through a domain wall, even if the magnons are scattered. Wherever a thermal gradient exists, so too does the SSE, so the thermal magnons can then be "re-excited" on the other side of the domain wall. This is starkly different from the electrically excited magnons, which, following scattering cannot be “re-excited.” As such, it is not surprising that only second harmonic switching data is observed, and the findings presented here support the conclusions of a growing body of work studying magnon scattering at domain walls.

## Section 2. Electrode spacing dependence.

Here we study the influence of electrode spacing on the observed thermal magnon signal. The range of spacings achievable in our experiment differs from that of systems, such as YIG, where the signal is modulated with an applied magnetic field. Owing to the need for electric field control, and the high voltages required to reach the same electric field value at increased electrode spacing, we are limited in the number of spacings achievable. Our results (**Fig. S4**) for the differential $V_{nl}$ ($V_{nl}$(positively poled) – $V_{nl}$(negatively poled)) for an electrode spacing series, fits well to an exponential decay function, as would be expected in a 1D spin diffusion model [2].

As noted in [3] the distance dependence can be a function of both the distance dependence of the thermal gradient, $\nabla T(x)$, and magnon relaxation. Further, as in [12], magnon accumulation can occur at interfaces with variable spin opacities, so it is possible that there exists a cascading magnon accumulation at successive domain wall interfaces. These two effects combine to give the net thermal magnon current, observed. Considering a simple model of phonons as heat carriers, we expect diffusive phonon heat transport, and, similar to the model for diffusive magnon transport [2], we expect the resulting thermal gradient (from phonons alone) to exhibit exponential dependence on distance. Therefore, if we consider only the domains *directly beneath* the Pt detector (and their magnetic state, as dictated by the electric field control) as being relevant to the detected thermal magnon signal, we expect to record the observed exponential dependence. We stress that this is a limiting case where magnons are completely scattered any time they encounter a domain wall, the effects of magnon diffusive transport resulting from magnon accumulation at domain walls are ignored, and phonons are the dominant heat carrier. While it is likely that the contribution to magnon signal from domains directly adjacent to the Pt detector are significant (dominant), as seen in **Fig. S5** one must also consider magnons thermally excited magnons near the detector, which, although attenuated, do not completely scatter to zero amplitude when traversing a domain wall, or undergo diffusive transport following the magnon accumulation profile. Finally, one must consider that magnons also carry heat. It is possible, then, that the electrode spacing dependence contains two exponential decay terms, one from the thermal gradient (*i.e.*, phonons) and one from magnon relaxation. Disentangling these two contributions is the subject of ongoing research.

**Figure S4**. Electrode spacing series. Differential $V_{nl}$ as a function of electrode spacing on log-linear scale. Heater current is 800μA for all devices. Error bars represent standard deviation of differential $V_{nl}$ as measured by a lock-in amplifier for 200 seconds. Protocol is identical to **Main Text Fig 2**. Red dashed line shows fit to $y = A\,e^{-\frac{d}{\lambda}}$.

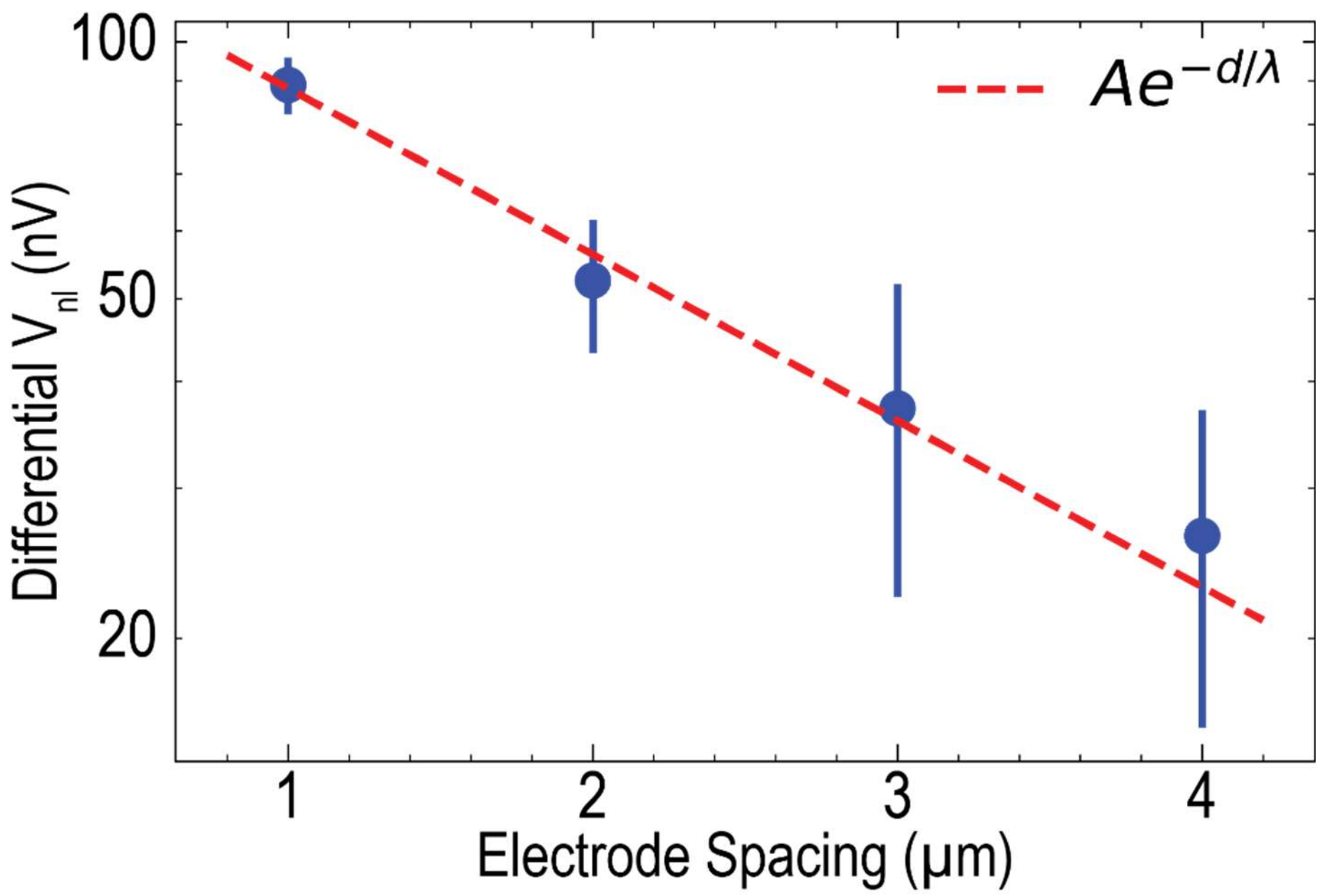

**Figure S5.** Schematic showing thermal magnon signal dominated by domains directly beneath the detector, as well as thermally activated magnons (notated by Red arrows) in nearby domains which are attenuated via scattering, though contribute a nonzero amplitude at the detector.

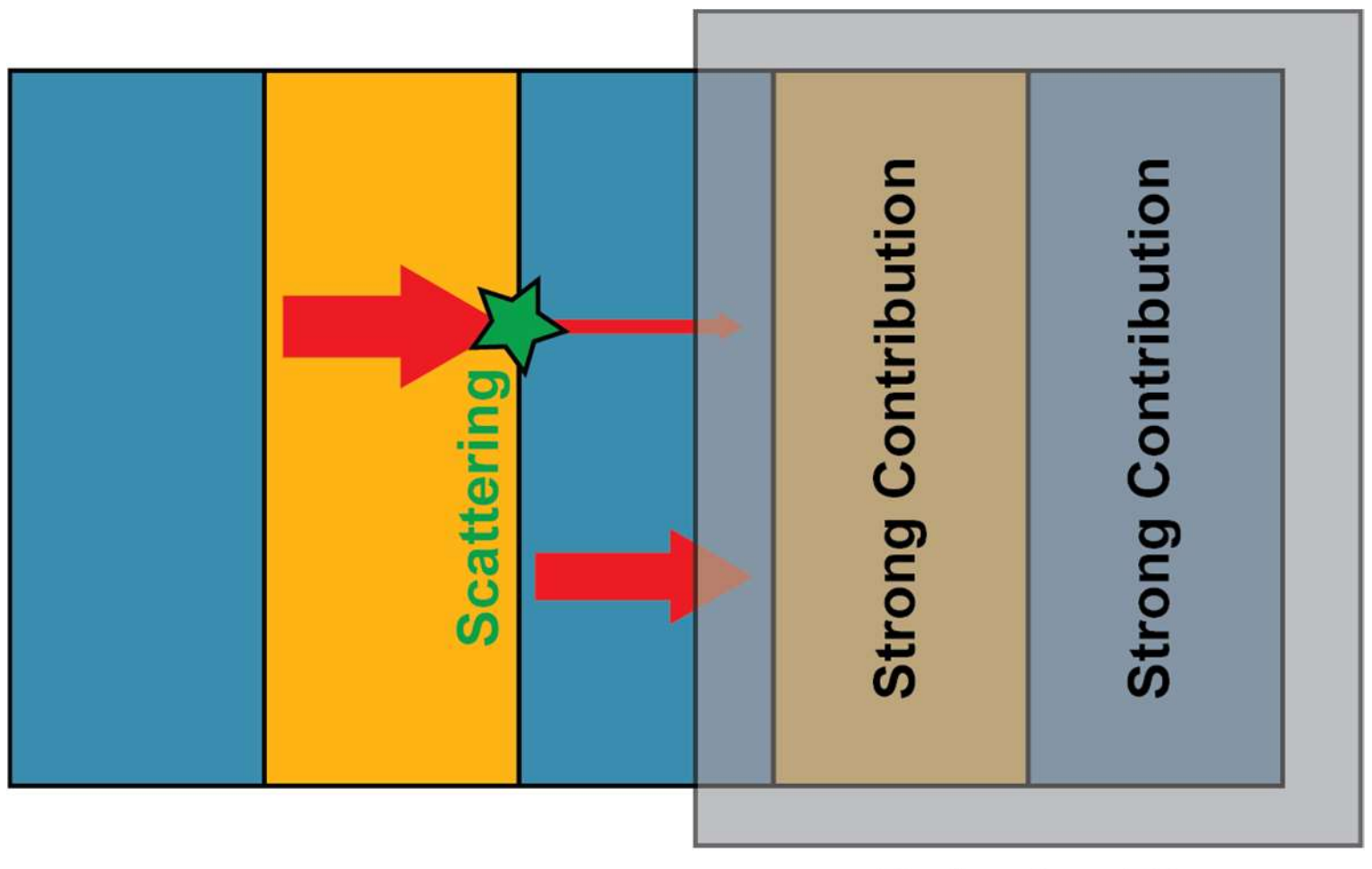

**Section 3. Magnetic Field Angular Dependence.**

Here, we address the question of magnetic-field dependence of the observed nonlocal signal. In typical nonlocal spin transport experiments, one rotates an applied magnetic field in the plane of the sample, while measuring both the first and second harmonic nonlocal voltages. With a sufficiently high magnetic field, the magnetic order (ferri-, ferro-, or antiferromagnetic) can be rotated along with the field. In-plane rotation of the applied field then results in a sinusoidal modulation of the observed non-local signal [5,13]. We show the results of in-plane magnetic field angular dependent measurements on BFO (**Supplementary Material Fig. S6**), which exhibit $\pi$ −periodic oscillation (linearly dependent on the applied magnetic field magnitude), apparently consistent with the SSE. Upon closer examination, however, it is revealed that this signal is dominated by the Nernst effect in the Pt. The Nernst effect is the thermal analog of the Hall effect in a metal, where an out-of-plane magnetic field and transverse thermal gradient produce a voltage proportional to the cross product of the two, *i.e.,* $V_{Nernst} \propto \boldsymbol{H} \times \boldsymbol{\nabla T}$. Careful study of the observed sinusoidal oscillation (**Supplementary Material Section 4,5**), shows that it can be attributed to a small misalignment during mounting creating a ~2.8 degree tilt out-of-plane. This then begs the question: where is the thermal magnon signal in the magnetic field dependence? The non-local SSE signal is very small, since the field is not high enough to have an impact on the magnetic order. Estimates of the spin flop magnetic field in BFO, for example, are as high as 15-20T [14], meaning that at the fields studied ($\leq 6$ T) the signal from the Nernst effect completely dominates any potential signal from the SSE, which is known to be weak below the spin flop transition [15]. Furthermore, the Nernst effect signal will scale linearly with the applied magnetic field magnitude, making it difficult to distinguish the magnon signal from the Nernst effect signal. In order to properly do so, one would ideally have multi-axis control of the applied magnetic field direction and could minimize (correct for out-of-plane tilting) the Nernst effect at low field, before ramping above the spin flop field.

While the magnetic field has no measurable effect on the magnonic nonlocal signal, we can still observe electric field control in the magnetic field angular dependent measurements. Upon electric field poling, we can induce an offset in the sinusoidally oscillating Nernst effect signal (**Supplementary Material Fig. S7**). The magnitude of this nonlocal voltage offset, after correcting for variability in resistance across different devices (on both the detector and injector Pt wires), matches that predicted by the quadratic scaling nonlocal current on heater power (**Main Fig. 4c**).

This is a noteworthy finding, indicating that the electric field control is not only remnant, but robust to spuriously applied magnetic fields.

**Figure S6. Second Harmonic Voltage for nominal in-plane and out-of-plane orientations.** Left data shows nominal in-plane mounting of the sample with $\pi$ −periodic oscillation of second harmonic voltage on the detector wire (consistent with SSE or Nernst effect). Right panel shows out-of-plane mounting and linear dependence of measured second harmonic voltage (consistent with the Nernst effect). Both right and left are measured with the same heater current. These measurements (see **Supplemental Material Section 4,5**) indicate that just ~2.8 degrees of out-of-plane tilt from mounting the sample "in-plane" can result in the signal observed in the left panel, even in the absence of SSE.

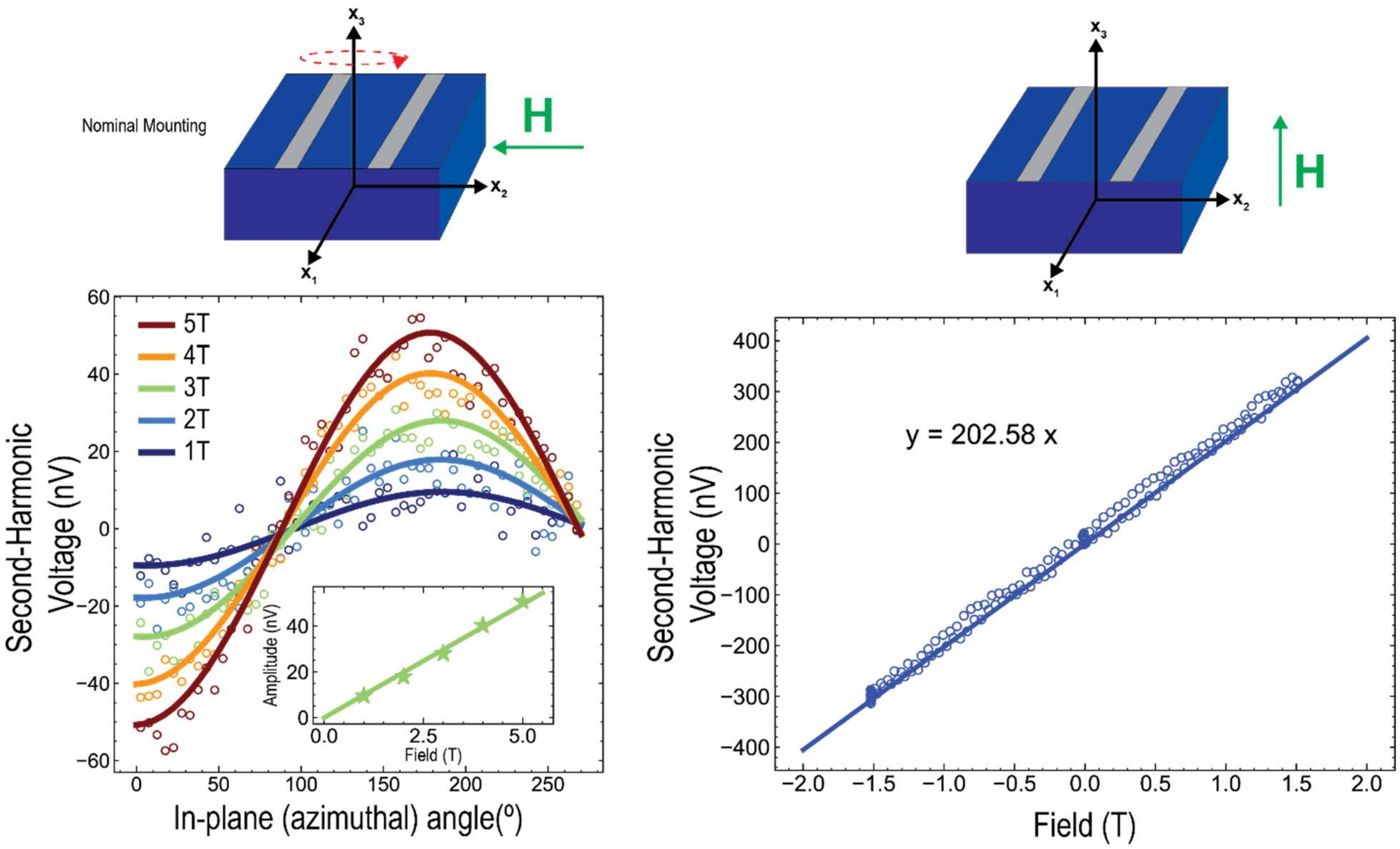

## Section 4. Calculation of Nernst Effect Signal Resulting from small out-of-plane tilt

The Nernst effect in a solid is the emergence of an electric field, $\boldsymbol{E}$, orthogonal to an applied magnetic field, $\boldsymbol{H}$, and a temperature gradient, $\nabla T$, in the solid, *i.e.,* $\boldsymbol{E_N} \propto \nabla \mathrm{T} \times \mathbf{H}$. A non-local geometry is highly susceptible to Nernst effect signals, whereby, a material mounted with a slight tilt with respect to an otherwise "in-plane" magnetic field (**Supplementary Material Fig. S8**) can display a signal with identical qualitatively identical features to that of a non-local spin Seebeck signal. We derive this in detail in the following:

Let the applied magnetic field direction be along $x_2$. We first solve the case of no tilt. $x_1', x_2', x_3'$ describe coordinate axes rotating with the sample (around $x_3$). The thermal gradient $\nabla T$ points across the channel, *i.e.* $\nabla \hat{T} \propto \widehat{x_2'}$. The measurement is sensitive only to electric field (in the form of a voltage) along the wire, *i.e.,* $V \propto E_{x_1'}$. We solve for the electric field from the Nernst effect:

$$\boldsymbol{E} \propto \nabla \hat{T} \times \boldsymbol{H} \propto \widehat{x_2'} \times \widehat{x_2} \propto (R_3(\phi)\widehat{x_2}) \times \widehat{x_2}$$

where $R_z(\phi)$ is the rotation matrix for a rotation $\phi$ around $x_3$, *i.e.*

$$R_3 = \begin{pmatrix} \cos\phi & -\sin\phi & 0 \\ \sin\phi & \cos\phi & 0 \\ 0 & 0 & 1 \end{pmatrix}$$

$$\rightarrow \boldsymbol{E} \propto (-\sin\phi, \cos\phi, 0) \times (0,1,0) = (0,0,-\sin\phi)$$

Indicating that there is no detected non-local voltage ($E_{x_1'} = 0$) for the case of no out-of-plane tilt. Next we solve for the case with a small tilt angle (meaning the sample does not lie perfectly flat in the plane in which the external magnetic field is applied). Without loss of generality, let this tilt be parameterized by an angle $\theta$ and occur via a rotation about $x_1$, *i.e.* the tilted sample before rotating about $x_3$ has coordinate axes:

$$\widehat{x_1'} = R_1(\theta)\widehat{x_1}$$

$$\widehat{x_2'} = R_1(\theta)\widehat{x_2}$$

$$\widehat{x_3'} = R_1(\theta)\widehat{x_3}$$

where

$$R_1 = \begin{pmatrix} 1 & 0 & 0 \\ 0 & \cos\theta & -\sin\theta \\ 0 & \sin\theta & \cos\theta \end{pmatrix}$$

now, upon rotating the sample around $x_3$, we solve for the induced electric field from the Nernst effect.

$$\boldsymbol{E} \propto \nabla \hat{T} \times \boldsymbol{H} \propto \widehat{x_2'} \times \widehat{x_2} = (R_3(\phi)R_1(\theta)\widehat{x_2}) \times \widehat{x_2}$$

$$= (-\cos\theta\sin\phi, \cos\theta\sin\phi, \sin\theta) \times (0,1,0) = (-\sin\theta, 0, -\cos\theta\sin\phi)$$

and the measured voltage along the detector wire is then

$$V \propto \boldsymbol{E} \cdot \widehat{x_1'} = -\,sin\,\theta\;\; cos\,\phi$$

which shows π-periodic dependence on the azimuthal (in-plane) angle, ϕ, precisely the same dependence one would expect from the nonlocal voltage. Furthermore, the thermal gradient scales with the square of the applied current, meaning that Nernst signals and non-local SSE signals show qualitatively *the same* current and angle dependence. There are a few key signatures of Nernst signals.

1. Linear dependence on applied magnetic field. While this may come from lack of saturation of the magnetic order in the material (especially the case in antiferromagnets), linear dependence of applied field is consistent with the Nernst effect.
2. Variability in phase and or amplitude of the signal when remounting the sample (**Supplementary Material Fig. S9**). Each time one remounts the sample, in principle one introduces a random tilt, which will result in a random amplitude (random θ) and random phase (as we tilts can occur from rotations along either $x_1$ or $x_2$.
3. Orthogonally fabricated devices give the same angle dependence.

To ensure that a measured signal is not from the Nernst effect, one should mount the sample normal to the applied magnetic field, and measure a field dependence of the measured voltage on the detector wire for fixed current on the heater wire. In this case, the measured Nernst voltage is linearly dependent on field and given by:

$$V = \alpha H j^2$$

Where $H$ is the applied magnetic field amplitude, *j* is the heater current amplitude, and α is a fitting parameter, which is fit to the experimentally obtained data. Returning now to the “in-plane” angular dependence, given the amplitude of the sinusoidally varying signal, A, one can calculate the required erroneous out-of-plane tilt angle to produce such a signal as:

$$\theta_{\mathrm{tilt}} = \arcsin(A/j^2 \alpha \mathrm{H})$$

using the $\alpha$ measured in the OOP configuration.

As observed (**Supplementary Material Fig. S6**), this analysis performed on our BFO sample revealed a required tilt of just ~2.8 degrees. As a rule of thumb, mounting of the sample introduces <5 degree out-of-plane tilt. If the required tilt is very large, then it is unlikely that your signal stems *solely* from the Nernst effect.

This ignores that one will undoubtedly introduce a small tilt in the out-of-plane mounting as well. If one so desires, the sample can be mounted several times in order to obtain several values for $\alpha$. The researcher should then use the maximum value of $\alpha$ obtained across remount trials.

If one has access to multi-axis alignment, it behooves one to rotate in-plane to a maximum of the $\pi-$periodic signal, and minimize the measured voltage using other axes of alignment. This will help correct for out-of-plane tilts.

**Section 5. Calculation of tilt degree causing sinusoidal signal (Nernst effect).**

Following the discussion if **Section Material Section 4**, we fit the out-of-plane magnetic field dependence to the equation:

$$V = \alpha H j^2$$

Using our results (**Supplementary Material Fig. S6**) we can extract a value of $j^2\alpha = (202.58\mathrm{nV/T})$. Using the observed amplitude (**Supplementary Material Fig. S6**) of the sinusoidally varying signal at 5T, we find:

$$\theta_{\mathrm{tilt}} = \arcsin\left(\frac{(50\mathrm{nV})}{(202.58\mathrm{nV/T})(5\mathrm{T})}\right) = \arcsin\left(\frac{10}{202.58}\right) \approx 2.8^{\circ}$$

Such a finding confirms that the magnetic field angular dependence of the second harmonic voltage in the detector wire is dominated by the Nernst effect. However, this does not contradict the electric-field control in the absence of an applied magnetic field. Further, as shown by poling the sample with an electric field, then measuring the angular dependent Nernst effect signal (**Supplementary Material Fig. S7**), one observes an offset between the two poled states, indicating that not only is the electric-field poling non-volatile, but in fact robust to externally applied magnetic fields!

**Figure S7. Non-volatility of Electric-field control; robust to applied magnetic fields.** Nernst effect (angle dependent) signal before and after electric field poling. The data reveal that while the Nernst effect dominates the angular dependence with applied *magnetic* field, *electric* field poling induces non-volatile changes to the magnetic order which are robust to externally applied magnetic fields (consistent with the exceptionally high spin flop field in $BiFeO_3$). Data shown is at 5T applied magnetic field.

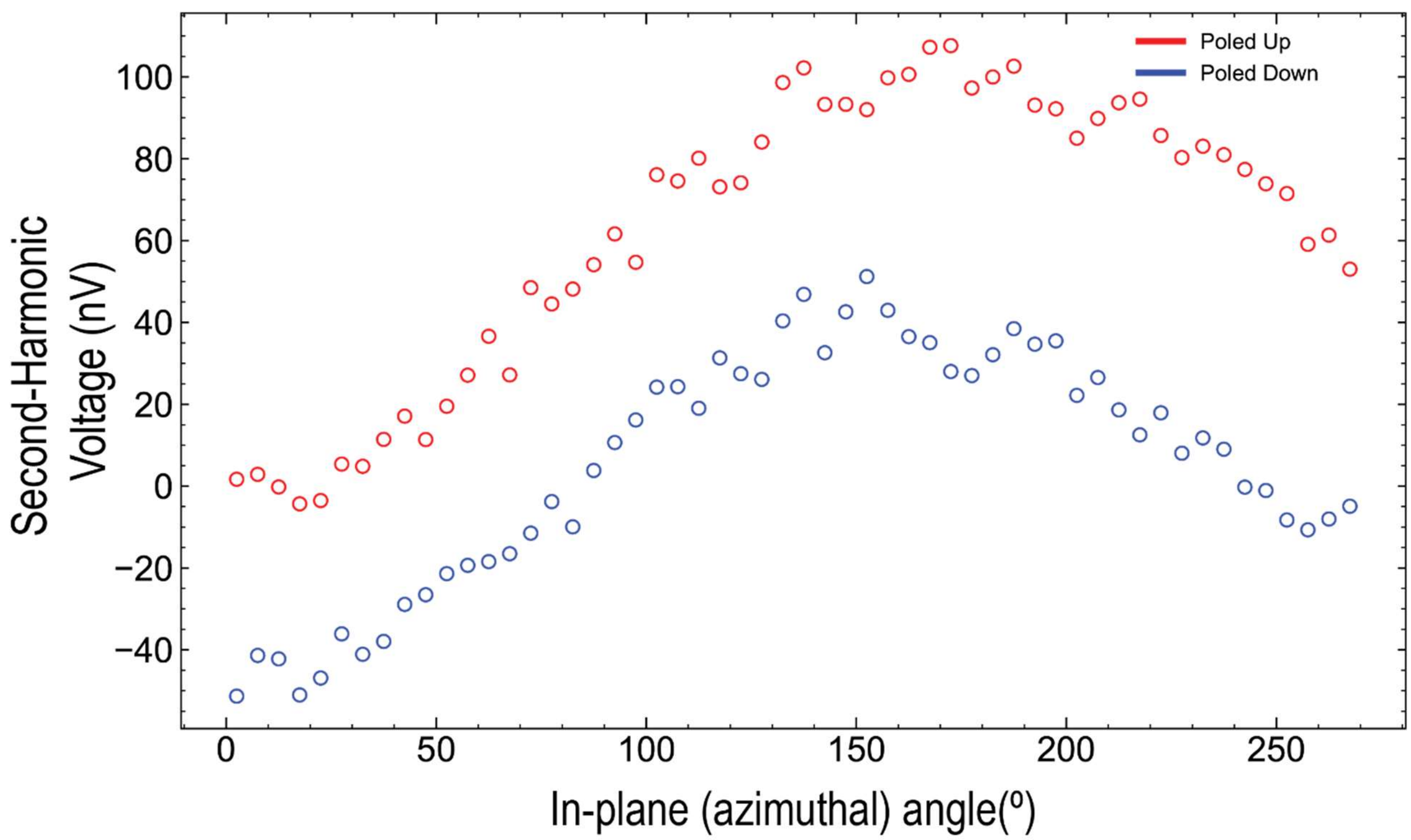

**Figure S8. Out-of-plane tilt during mounting. (Nernst Effect).** A small tilt when mounting the sample can lead to erroneous Nernst effect signals which can qualitatively match the expected non-local signal from the SSE. See Supplementary Section X for a detailed analysis.

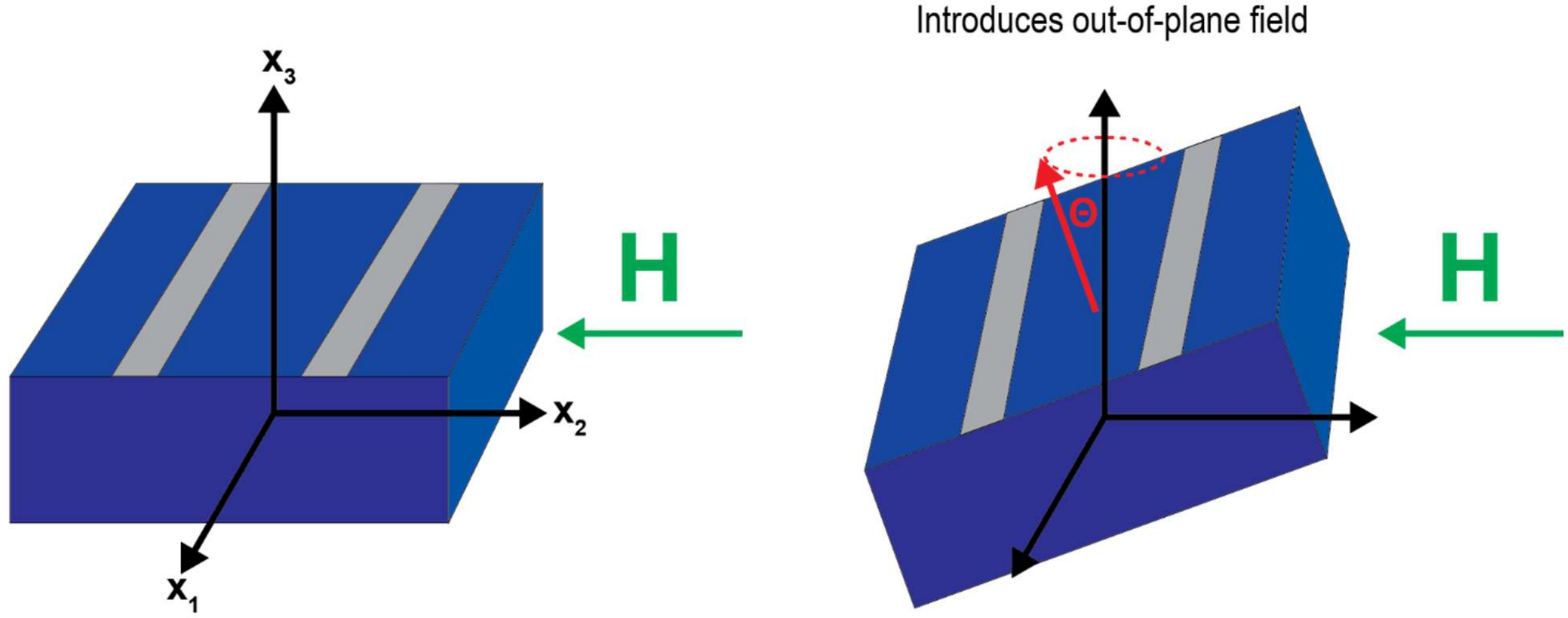

**Figure S9. Magnetic field angular dependence after remount.** If Nernst effects dominate the (nominal) “in-plane” mounted signal, remounting of the sample is predicted to cause random phase and amplitude offsets of the angle dependence. Indeed this is observed here. In both Blue (before remount) and Red (after remount) curves, a constant offset is subtracted.

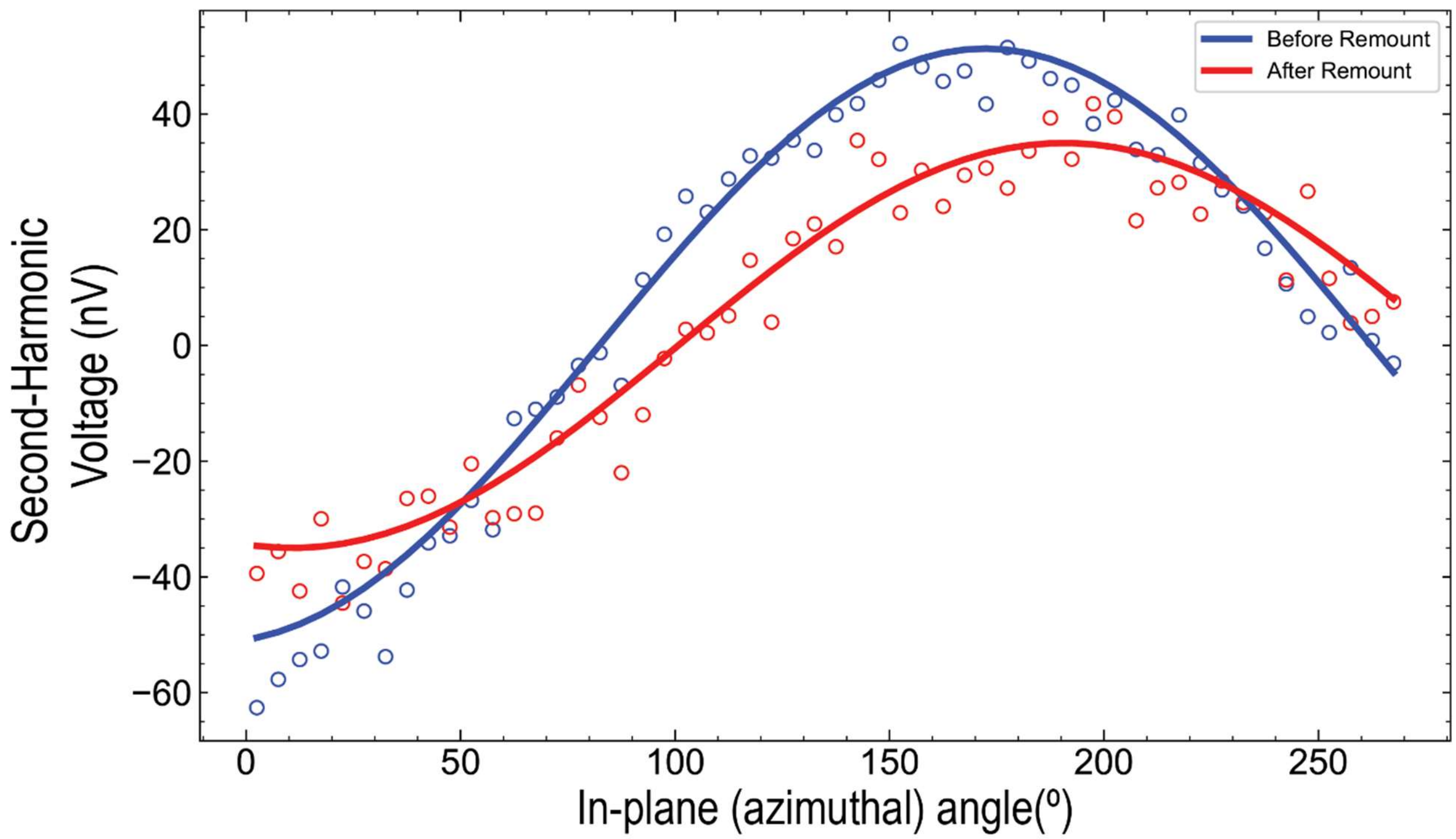